\documentclass[twocolumn,showpacs,preprintnumbers,amsmath,amssymb]{revtex4}

\usepackage{color}

\usepackage{graphics}
\usepackage{graphicx}
\usepackage{dcolumn}
\usepackage{bm}
\usepackage{amssymb}
\begin{document}
\preprint{APS/123-QED}

\title{Microwave dielectric study of spin-Peierls and charge ordering transitions in
(TMTTF)$_2$PF$_6$ salts}
\author{Alexandre Langlois}
\author{Mario Poirier}
\author{Claude Bourbonnais}
\affiliation{Regroupement Qu\'eb\'ecois sur les Mat\'eriaux de
Pointe, D\'epartement de Physique, Universit\'e de Sherbrooke,
Sherbrooke, Qu\'ebec,Canada J1K 2R1}
\author{Pascale Foury-Leylekian}
\author{Alec Moradpour}
\author{Jean-Paul Pouget}
\affiliation{ Laboratoire de Physique des Solides, CNRS UMR 8502,
Universit\'e Paris-Sud, 91405 Orsay C\'edex, France}

\date{\today}

\begin{abstract}
We report a study of the 16.5 GHz dielectric function of
 hydrogenated   and deuterated organic salts (TMTTF)$_2$PF$_6$. The
temperature behavior of the dielectric function is consistent with
short-range polar order whose relaxation time decreases rapidly
below the charge ordering temperature. If this transition has
more a relaxor character in the  hydrogenated  salt, charge
ordering is strengthened in the deuterated one where the
transition temperature has increased by more than thirty percent.
Anomalies in  the dielectric function are also observed in the
spin-Peierls ground state revealing some  intricate lattice
effects in a temperature range where both phases coexist.    The
variation of the  spin-Peierls ordering  temperature under
magnetic field appears to follow a mean-field prediction despite
the presence of spin-Peierls fluctuations over a very wide
temperature range in the charge ordered state of these salts.

\end{abstract}

\pacs{71.20.Rv,71.30.+h,77.22.Ej}
\maketitle
\section{INTRODUCTION}
The Fabre ((TMTTF)$_2$X) and Bechgaard ((TMTSF)$_2$X) series of
charge transfer salts  show a very rich sequence of competing
ground states when either hydrostatic or chemical pressure is
applied \cite{Bourbonnais2008}.  In the universal   phase diagram
of these series, the (TMTTF)$_2$X are Mott insulators that
develop a charge ordered (CO) state \cite{Brown2008,Monceau2008}.
This state is followed at lower temperature by either an
antiferromagnetic N\'eel  or a lattice distorted spin-Peierls
(SP) phase. This pattern turns out to be affected by  pressure,
anion X  substitution, and to some degree by   deuteration of the
methyl groups. This is how the (TMTTF)$_2$X  can be   moved along
the pressure axis with respect  to the  (TMTSF)$_2$X series for
which metallic, antiferromagnetic and superconducting phases can
be stabilized. In low pressure conditions, the
quasi-one-dimensional (TMTTF)$_2$X   salts thus appear as model
correlated systems to study the  interplay between spin, charge,
and lattice degrees of freedom.

In several (TMTTF)$_2$X salts with octahedral  anions X, the CO
transition follows and in some cases coincides with  \ the
4\textit{$k_F$} charge localization
\cite{Coulon1985,Javadi1988,Nad99,Chow2000,Monceau2001}. The
first evidences of this transition in X= SbF$_6$ and AsF$_6$ came
from transport measurements
\cite{Coulon1985,Javadi1988,Laversanne1984}. The transition was
dubbed ``structureless'', because of the absence of any
structural modification associated to it \cite{Laversanne1984}.
The CO character of the transition only comes much later from the
low frequency dielectric response\ \cite{Nad99,Nad2000}, and NMR
experiments in which charge disproportionation in the unit cell
was  unveiled \cite{Chow2000,Zamborsky2002,Fujiyama2006}. The CO
character of the transition  is also found from  infrared
spectroscopy measurements \cite{Dumm2006}. The CO transition is
also accompanied by the onset of a ferroelectric state revealed
by the divergence of the low frequency dielectric constant
\cite{Nad2006}.

As regards to spin degrees of freedom, these are essentially
decoupled  from  the progressive charge localization  or the CO
transition.  However, in compounds like X= PF$_6$ in normal
pressure conditions, a SP transition takes place at $T_{\rm SP}
\approx 18$~K (hydrogenated) \cite{Pouget1982,Foury04} and 13 K
(deuterated) \cite{Pouget2006}. The spin singlet state that goes
with the SP lattice distortion has been borne  out  by  spin
susceptibility \cite{Creuzet1987,Dumm00,Foury08}, and NMR spin
relaxation rate \cite{Creuzet1987,Wzietek1993,Zamborszky2002B},
whereas the magnetic field-temperature phase diagram for the
PF$_6$ salt has been obtained by NMR and high-field magnetization
studies \cite{Brown1998}.  In the PF$_6$ salt, the observation of
X-ray 2\textit{$k_F$} diffuse scattering   indicates the presence
of lattice  precursors of the SP transition below 60~K
\cite{Pouget1982}; these open a spin pseudo gap
\cite{Bourbonnais96},  as  exhibited by magnetic suceptibility
and NMR data \cite{Creuzet1985,Creuzet1987,Foury08}.

Finally, lattice  expansion  effects were recently observed at
both CO and SP transitions of the PF$_6$ and AsF$_6$  salts
\cite{deSouza2008,deSouza2009}. Indeed, distinct lattice effects
were found at $T_{\rm CO}$ in the uniaxial expansivity along the
interstack $c^*$ direction, signaling an  active role of anion X
lattice degrees of freedom  in the  stabilization of the CO
ground state, and further clarifying the ferroelectric nature of
the transition. It is also the $c^*$-axis expansivity that is the
most strongly affected at the SP transition.

In this paper, we address the issue of interplay between charge,
spin and lattice degrees of freedom by studying the 16.5 GHz
complex microwave dielectric function of (TMTTF)$_2$PF$_6$ single
crystals  under magnetic field.  Dielectric anomalies are
observed  at both the CO  and SP transitions. The increase  of
the $T_{\rm CO}$ scale and the drop of $T_{\rm SP}$ are also
confirmed for the deuterated compound as a negative shift on the
pressure scale. Up to 18 Tesla, the SP transition temperature
varies quadratically with  field,  in fair agreement with  the
mean-field prediction obtained by  Cross \cite{Cross1979}. Other
magnetic fields effects on the polarizability of the system are
found and ascribed to the interaction between CO and SP
orderings. Finally, we report the existence  of thermal
relaxation effects in the deuterated compound near the SP
transition.

\section{EXPERIMENT}
Hydrogenated (H{\tiny{12}}) (TMTTF)$_2$PF$_6$ single crystals were
grown from THF using the standard constant-current (low current
density) electrochemical procedure. The synthesis of the fully
deuterated TMTTF-$d_{12}$ (D{\tiny{12}}) was attempted using the
procedure of Wudl \textit{et al.} \cite{Wudl1981} for the
preparation of 3-chloro-2-butanone-d$_7$. However, this awkward
very poorly described procedure led in our hands to traces, if
any, of the desired chloroketone. Therefore, we worked out a more
practical and shortened procedure to prepare this compound:
perdeuterioacetoin was first prepared from perdeuteriobiacetyl
\cite{During1971} according to the literature \cite{Pechmann1980}
and the procedure for the conversion of allylic alcohols into
chlorides \cite{Magid1981} applied to this -hydroxy ketone. To
the cooled mixture of the latter with a slight excess of PPh$_3$
in sulfolane was added dropwise a solution of hexachloroacetone in
sulfolane, and the reaction mixture stirred at room temperature
for two weeks; a flash distillation of the reaction mixture and a
subsequent distillation at atmospheric pressure led to the desired
chloroketone with a (up to) 50\% yield. Starting from this
deuterated  chloroketone, TMTTF-$d_{12}$ was obtained using usual
procedures and the corresponding deuterium incorporation-levels
estimated by NMR to be 97.5 \%.

The crystals have the shape of a needle oriented along the chain
axis $\textit{\textbf{a}}$, from which were cut short slabs of
typical dimensions 2.4x0.8x0.4 mm$^3$ for H{\tiny{12}} and
1.5x0.4x0.07 mm$^3$ for D{\tiny{12}}. We used a standard
microwave cavity perturbation technique \cite{BURAVOV} to measure
the complex dielectric function $\epsilon^* = {\epsilon}^{'} + i
{\epsilon}^{''}$ along the $\textit{\textbf{a}}$ axis.  A copper
cavity resonating in the TE$_{102}$ mode was used at 16.5 GHz.
The organic slab is inserted in a mylar envelope and immobilized
by thin cotton threads to prevent any stress or movement during
thermal cycling. The envelope is glued on a quartz rod to allow
its insertion in the cavity and the precise orientation of the
slab along the microwave electric field. Following the insertion
of the sample, changes in the relative complex resonance
frequency $\Delta f/f + i \Delta(1/2Q)$ ($Q$ is the cavity
quality factor) as a function of temperature are treated
according to the depolarization regime analysis after
substraction of the envelope contribution. With this microwave
technique, the slab is associated to a prolate ellipsoid and
absolute values can only be measured within 30\% of accuracy. A
magnetic field up to 18 Tesla could be applied along the
$\textit{\textbf{c$^*$}}$ axis.

\begin{figure}[H,h]
\includegraphics[width=8.5cm]{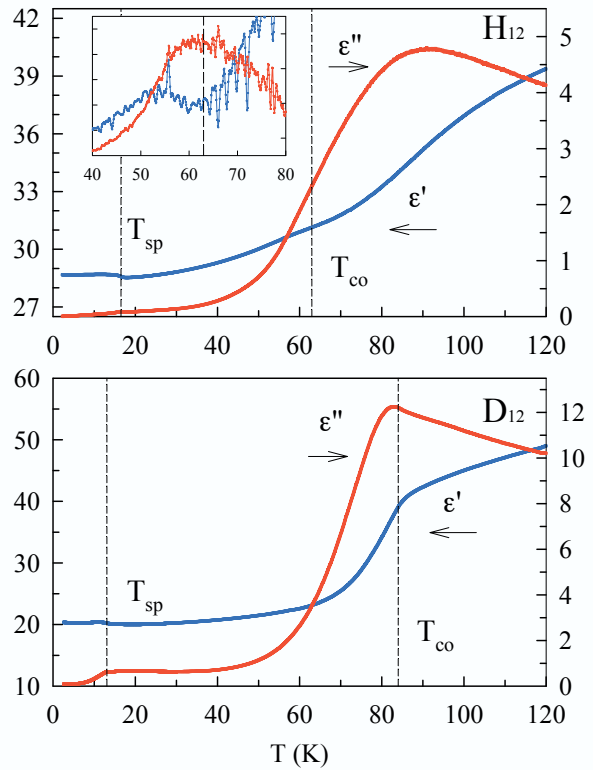} \caption{(Color online)
Temperature dependence of the dielectric function along the chain
axis at 16.5 GHz: top panel PF$_6$(H{\tiny{12}}), definition of
$T_{\rm CO}$ from the derivatives shown in the inset; lower panel
PF$_6$({D\tiny{12}}).}\label{fig.1}
\end{figure}

\section{RESULTS AND DISCUSSION}

When the temperature is decreased from 300 K, the dielectric
function of both H{\tiny{12}} and D{\tiny{12}} salts shows a
similar behavior: the real part $\epsilon'$ decreases smoothly
with a faster rate below 100 K before  its saturation at low
temperatures; the imaginary part $\epsilon''$ rather increases
first, reaches  a maximum below 100 K and decreases rapidly
toward zero at low $T$. These temperature profiles are presented
below 120 K in Fig.~\ref{fig.1}. The features observed below 100
K are sharper for the D{\tiny{12}} compound and they are
attributed to a well-defined CO transition at $T_{\rm CO}$ = 84 K
as it is clearly identified  for both parts of the complex
$\epsilon^*$; this value of $T_{\rm CO}$ is identical to the one
deduced from ESR measurements \cite{Coulon2007}. The
determination of $T_{\rm CO}$ is more difficult for the
H{\tiny{12}} salt, because of the diffusive character of the
transition identified of the relaxor type \cite{Nad99,Nad2000}.
By looking at the temperature derivative of both parts of the
dielectric function (see inset of Fig.~\ref{fig.1}, top
panel),    we deduce the value $T_{\rm CO}$ $\simeq$ 63 K,   in
agreement with previous works
 \cite{Chow2000}. Contrary to previous dielectric
measurements on the same compound \cite{Nad2000},  however, our
microwave data clearly show anomalies in the low temperature
range at $T_{\rm SP}$, which will be analyzed later in the paper.

\begin{figure}[H,h]
\includegraphics[width=8.5cm]{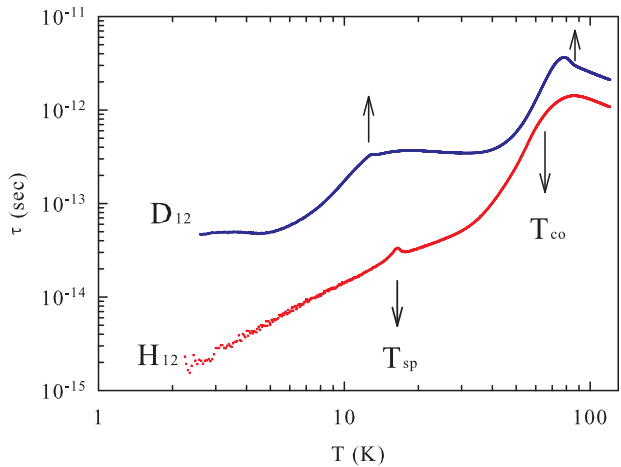} \caption{(Color online)
Temperature dependence of the relaxation time $\tau(T)$ obtained
from Eq.~\ref{eq.1}.} \label{fig.2}
\end{figure}

Earlier analysis of the dielectric response measurements
performed at   lower frequencies (10$^3$-10$^7$ Hz)
\cite{Nad2000,Nad2006}, showed that the $\epsilon'(T)$ data can
be qualitatively understood by assuming the existence of critical
slowing down near a phase transition. The maximum of
$\epsilon''(T)$ was used to obtain an effective temperature
dependent relaxation time $\tau(T)$. In a similar way, we propose
to model the frequency dependence of the dielectric function by
using a monodispersive Debye relaxation term to get the
temperature dependence of the relaxation time at a single
frequency $\omega$ from the ratio
\begin{equation}
\tau(T) = \frac{\epsilon''(T)}{\omega [\epsilon'(T)-
{\epsilon}_\infty]}\label{eq.1}
\end{equation}
The temperature profile $\tau(T)$ deduced from Eq.~\ref{eq.1} is
not very sensitive to the exact value chosen for the high
frequency dielectric constant $\epsilon_\infty$, which was fixed
to 2.5 \cite{Jacobsen1983}. The $\tau(T)$ curves deduced for both
salts are shown in Fig.~\ref{fig.2} on a log-log scale. These
relaxation time curves are quite different from  the data
collected  at lower frequencies \cite{Nad2000}: not only are the
absolute values smaller by 4 orders of magnitude at 100 K, but the
temperature variation is   the  opposite below the CO transition.
Indeed, for the H{\tiny{12}} salt, although $\tau(T)$ increases
below 120 K, it presents a maximum around 80 K and decreases
rapidly by more than one order of magnitude down to 40 K.  Below,
$\tau(T)$ keeps decreasing, but with a different slope.  Between
80 K  and the lowest temperature reached at 2 K,  $\tau(T)$, has
practically decreased by 3 orders of magnitude,  while  the low
frequency data rather show an increase of the same size with only
a change of  slope  near $T_{\rm CO}$ \cite{Nad2000}.  A kink is
observed at $T_{\rm SP}$ followed by a small variation of the
decreasing rate. If one looks at  the relaxation time for the
D{\tiny{12}} salt in Fig.~\ref{fig.2}, it shows a similar
behavior from 120 to 40 K, but the maximum in $\tau(T)$, which is
sharper, occurs just below $T_{\rm CO}$. For $T < $ 40 K, the
decrease is less pronounced than for the H{\tiny{12}} salt due to
an additional contribution to the dielectric function, and which
will be discussed in more detail below.

\begin{figure}[H,h]
\includegraphics[width=8.5cm]{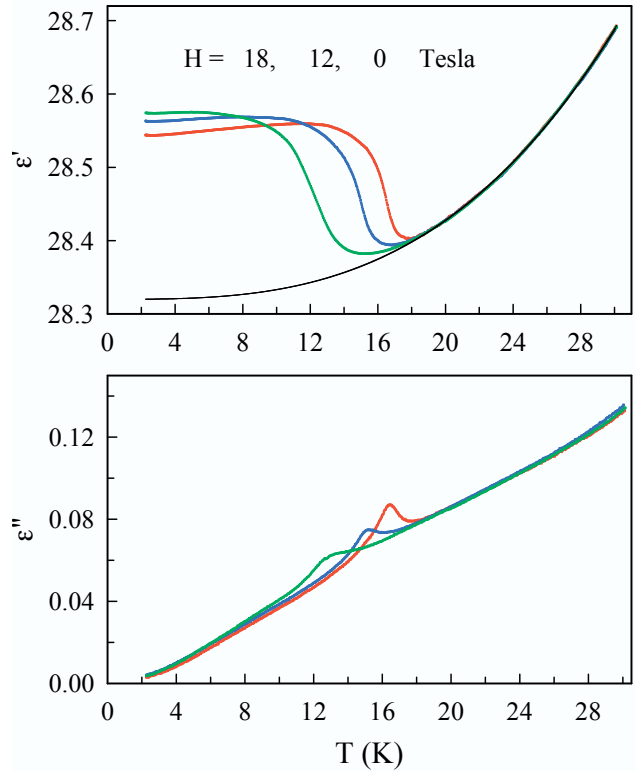} \caption{(Color online)
Temperature dependence of the the dielectric function near
$T_{\rm SP}$ for the H{\tiny{12}} salt in 0 (red), 12 (blue) and
18 (green) Tesla. The black line is the $a$ + $bT^3$ fit.}
\label{fig.3}
\end{figure}

The discrepancy between  microwave and low frequency data for the
H{\tiny{12}} salt may be  linked to  the nature of the relaxor
ferroelectric state and to the four orders of magnitude
difference in frequency. The microwave experiment is likely to
be  sensitive to the spatial variation of  polar order taking
place at relatively short length scale, which is characterized by
a short relaxation time. As we move further  below $T_{\rm CO}$
on the temperature axis, these polar fluctuations decrease in
size yielding a decrease of the dielectric constant and of the
corresponding losses, as depicted  in Figures~\ref{fig.1} and
\ref{fig.3}.

As regards to the increase of $T_{\rm CO}$ by deuteration,  it
has been shown recently that the collective displacement of
counteranions  X$^-$ is directly coupled to the charge modulation
along the stacks \cite{deSouza2008} and that the stabilizing
potential grows not only with the size of the anion but also upon
deuteration \cite{Pouget2006}. The increase of $T_{\rm CO}$
revealed by the microwave experiment on a deuterated crystal is
consistent with this picture.

\begin{figure}[H,h]
\includegraphics[width=8.5cm]{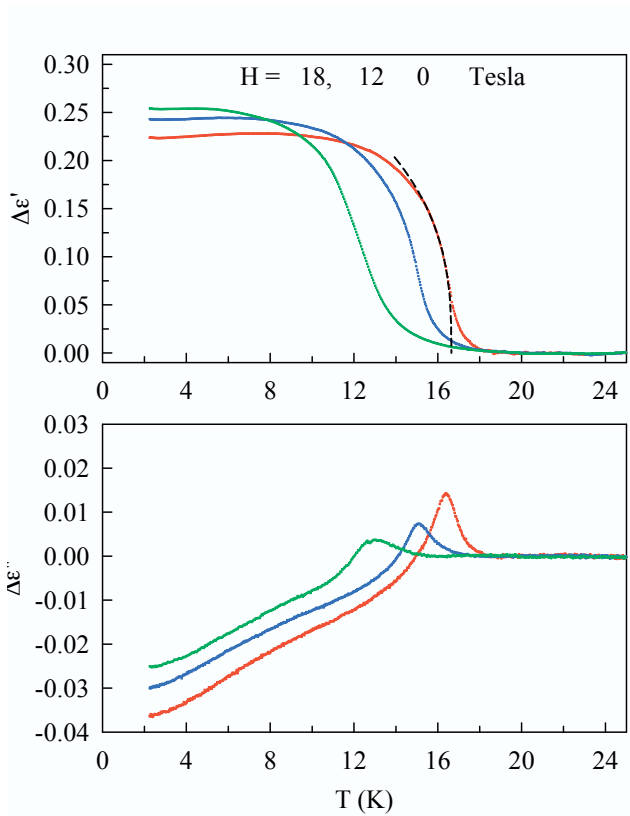} \caption{(Color online)
Contribution of the SP state to the dielectric function as a
function of temperature in 0, 12 and 18 Tesla field. The dashed
line is a fit to $\Delta \epsilon' \propto (T_{\rm SP}-T)^\beta$
($\beta \simeq 0.36$) in the critical region of the SP
transition.} \label{fig.4}
\end{figure}

Let us examine now the anomalies appearing in the SP state. We
show in Fig.~\ref{fig.3} the dielectric function of the
H{\tiny{12}} salt below 30 K. In zero magnetic field, the real
part $\epsilon'$ increases abruptly below 16.5 $\pm$ 0.1 K  and
then decreases slightly  as we move sufficiently down in
temperature. To the increase in $\epsilon'$ corresponds a peak
in  the imaginary part $\epsilon''$. The latter decreases to zero
with a larger slope. These anomalies are consistent with the
onset of the SP long-range order   at $T_{\rm SP} \simeq 16.5$K
\cite{Pouget2006}.  This is also compatible with   the
application of a magnetic field, which depresses  $T_{\rm SP}$
down to 12.5 K$\pm$ 0.2 at 18 Tesla . Except for a small
temperature interval where  critical  SP fluctuations enhance the
dielectric constant, the magnetic field has no  noticeable
effect  above $T_{\rm SP}$, where $\epsilon'$ follows a perfect
$T^3$ power law up to 30 K, as indicated by the black line
extrapolated toward   the very  low temperature domain  (top
panel). The microwave absorption is very low in this temperature
range and this yields larger imprecisions on $\epsilon''(T)$ when
the field is applied. Nevertheless, we found a $T^x$ power law
with $x$ lying in the 1.4-1.6 interval for $T > T_{\rm SP}$. We
have substracted these power laws extrapolated to zero to obtain
the contribution of the SP state to the dielectric function at
low temperatures, $\Delta\epsilon'$ and $\Delta\epsilon''$, shown
in Fig.~\ref{fig.4}.

The anomaly  shown by $\Delta\epsilon'$ as we dip into the SP
ordered region reveals an increase of polarizability in the
presence of the lattice distortion. This is compatible with a
reduction of the amplitude of the CO order parameter as the SP
order sets in, an effect also consistent with  previous  NMR  and
optical measurements on the cousin SP compound (TMTTF)$_2$AsF$_6$
\cite{Fujiyama2006,Dumm2006}. It is worth noting that the
amplitude of the variation shown  by  $\Delta \epsilon'$, as a
result of  SP ordering,   is at the most   1\% of the background
value exhibited in Fig.~\ref{fig.3}.   While the anomaly
indicates that the lattice distortion does alter the charge
ordered state of (TMTTF)$_2$PF$_6$, the relative smallness  of
the effect may explain the difficulty for  charge sensitive
probes  to detect  any reduction of  CO ordering by the SP state
in this material \cite{Dumm2006}.

The rapid increase of the dielectric response  below $T_{SP}$
suggests  a temperature dependence  of  $\Delta\epsilon'$ governed
by the criticality of the SP order parameter. A log-log plot
analysis of the $\Delta\epsilon'$ data reveals indeed that the
critical behavior can be reasonably  fitted  with a power law
$\Delta \epsilon' \propto (T_{\rm SP}-T)^\beta$. Using the value
$T_{\rm SP}$ obtained from the maximum slope of $ \Delta
\epsilon'$, the  exponent $\beta\simeq 0.36$ is extracted (dashed
line of Fig.~\ref{fig.4}). This non mean field value is consistent
with the $\beta$  expected for a SP -- one-component --  order
parameter in three dimensions.

Considering now the magnetic field dependence,
 $T_{\rm SP}$ is found to be progressively depressed under field. The
transition's width increases due to a larger temperature
interval  of SP critical fluctuations. However, at variance with
other spin-Peierls systems like CuGeO$_3$ \cite{Poirier1995}, or
Peierls systems \cite{Allen95},   for which the order parameter
is independent of magnetic field in the low temperature limit
\cite{Azzouz1996}, $\Delta\epsilon'( T\to0)$  is here field
dependent. The fact that for (TMTTF)$_2$PF$_6$, the SP lattice
distortion occurs in the  CO -- ferroelectric -- state may be
responsible for the slight increase of the low temperature
polarizability in the presence of a field.   As lattice effects
are involved in both transitions, it appears difficult, however,
to predict how both types of order  interact  and are modified by
a magnetic field.

\begin{figure}[H,h]
\includegraphics[width=8.5cm]{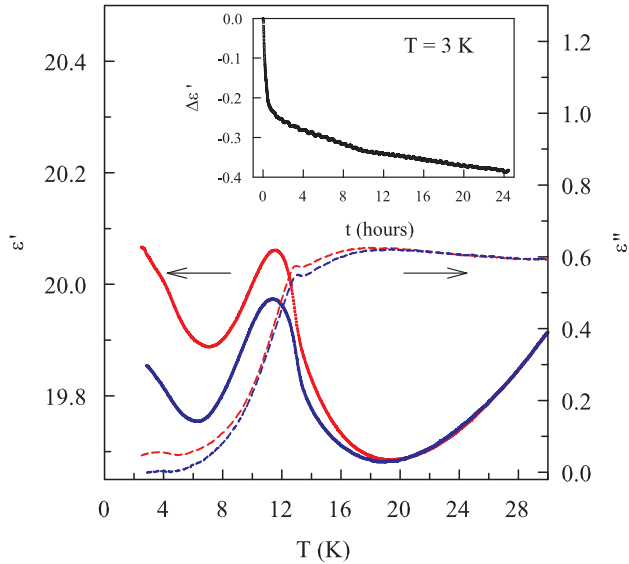} \caption{(Color
online) Temperature dependence of the dielectric function near
$T_{\rm SP}$ for the D{\tiny{12}} salt:  2 hours (red) and 24
hours (blue) waiting time at 3 K. Inset: time dependence of
$\Delta\epsilon'$ at 3 K.} \label{fig.5}
\end{figure}

As reported in neutron experiments \cite{Pouget2006}, we  also
observe a reduction of the  SP transition temperature  upon
deuteration. However,  thermal relaxation effects  are observed
at microwave frequencies at low temperatures for the D{\tiny{12}}
salt.  By deuteration the decrease of internal pressure increases
the volume of the anion cavity delimited by the methyl groups.
These deuterated groups can thus move more freely and are
probably the cause of the  anomalous relaxation effects.  The
dielectric function of this salt is shown below 30 K in
Fig.~\ref{fig.5}. The temperature dependence of both parts of
$\epsilon^*$ is quite different from the ones shown in
Fig.~\ref{fig.3} for the H{\tiny{12}} salt. Although we can
identify features related to the SP transition,   that is   a
sudden slope variation on $\epsilon'$ and a sharp peak on
$\epsilon''$ at 13.1 K, there is clearly another mechanism
contributing to the dielectric function in this temperature
range. This contribution is critically dependent on time as
evidenced by the two curves appearing in Fig.~\ref{fig.5}, taken
respectively 2 and 24 hours after the first cooling down to 2 K.
In the inset, we present the time dependence of $\Delta\epsilon'$
at $T$ = 3 K, just after cooling the sample from 40 K. Two time
scales are clearly observed   (insert of Fig.~\ref{fig.5}): a
relatively fast one during the first half-hour and a very slow
one for which saturation is obtained approximately after 48
hours. Similar effects with decreasing amplitude are observed up
to 40 K. These thermal relaxation effects related to deuterated
methyl groups  prevent  any analysis of the SP contribution to
the dielectric function similar to the one   given  in
Fig.~\ref{fig.4} for the hydrogenated samples.  These effects
also   mask the real temperature dependence of the relaxation
$\tau(T)$ below 40 K in the CO state (Fig.~\ref{fig.2}).
Nevertheless, our microwave data confirm the decrease  of $T_{\rm
SP}$ by roughly 30\% upon deuteration \cite{Pouget2006} and its
magnetic field dependence $T_{\rm SP}(H)$ could be studied up to
18 Tesla and compared with the
 hydrogenated  salt.

\begin{figure}[H,h]
\includegraphics[width=8.5cm]{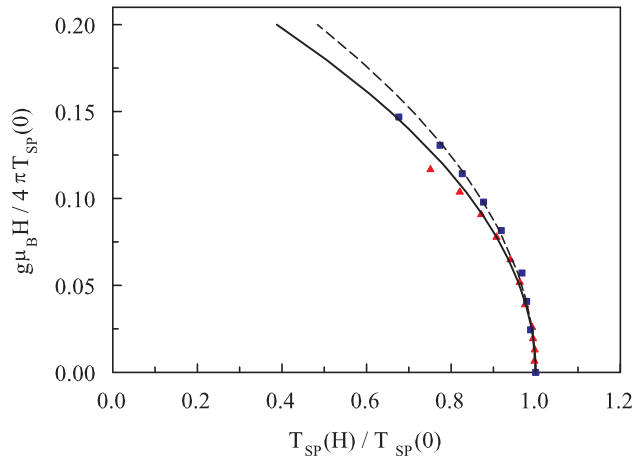} \caption{(Color
online) Magnetic field dependence of the reduced SP transition
temperature $T_{\rm SP}(H)/T_{\rm SP}(0)$ for the(TMTTF)$_2$PF$_6$
salts: red triangles for H{\tiny{12}}, blue squares for
D{\tiny{12}}; the continuous and dash lines are the respective
fits to Eq.~\ref{eq.2}.} \label{fig.6}
\end{figure}

The magnetic field dependence of the SP transition temperature is
known to fit  a quadratic variation \cite{Bloch1980,Poirier1995}
in relatively low fields as predicted theoretically
\cite{Cross1979,Azzouz1996}. In the hydrogenated
(TMTTF)$_2$PF$_6$ salt, NMR and  magnetization measurements have
indeed revealed  some  quadratic dependence \cite{Brown1998}, but
the entire range of low field values for $T_{\rm SP}(H)$ was not
available. From our field dependent microwave data, the phase
diagram including the D{\tiny{12}} and H{\tiny{12}} salts is
presented in Fig.~\ref{fig.6} and compared to the mean-field
prediction (continuous and dash lines) obtained by Cross
\cite{Cross1979}
\begin{equation}
\frac{T_{\rm SP}(H)}{T_{\rm SP}(0)} = \left[1 - c{\left(
\frac{g\mu_BH}{ 4\pi T_{\rm SP}(0)}\right)}^2\right],\label{eq.2}
\end{equation}
where the $g$-factor is fixed at 2 for (TMTTF)$_2$PF$_6$
\cite{Coulon1982}. For the H{\tiny{12}} salt, the fit  gives
$c\simeq  15.3 $, and $c\simeq  12.9 $ for the deuterated salt
which is  relatively close to the predicted value of $14.4$, in
comparison to other systems like CuGeO$_3$, where pronounced
deviations are found \cite{Zeman1999}.

\section{CONCLUSION}

In (TMTTF)$_2$PF$_6$ salts, we found that both the charge ordering
and  spin-Peierls transitions can be studied from measurements of
the microwave dielectric function. In the gigahertz frequency
range, the technique is sensitive to short-range polar order,
which yields a decrease of the dielectric constant below $T_{\rm
CO}$ due to a rapid reduction of these ordered regions and of
their relaxation time. The microwave data confirm not only the
relaxor character of the CO transition in the hydrogenated  salt,
but also its stabilization upon deuteration, which increases
$T_{\rm CO}$ by 33\%. Contrary to systems with no CO ordering,
dielectric anomalies are also observed in the SP state for both
salts, with a 20\% reduction of $T_{\rm SP}$ upon deuteration.
The opposite effects of deuteration on $T_{\rm SP}$ and $T_{\rm
CO}$ are compatible with an effective negative shift on the
pressure axis.

Thermal relaxation effects have been detected in the deuterated
salt, which have impeded a complete analysis of the SP ordering.
These SP anomalies confirm an indirect coupling between charge,
lattice and spin degrees of freedom due to intricate lattice
effects in a temperature range where both the SP and CO ground
states coexist. A quadratic magnetic field dependence of the SP
transition temperature appears to follow the mean-field
prediction  despite the existence of one-dimensional lattice
fluctuations  over a wide range of temperature in
(TMTTF)$_2$PF$_6$.

\acknowledgments{The authors achnowledge the technical support of
Mario Castonguay. This work was supported by grants from the Fonds
Qu\'eb\'ecois de la Recherche sur la Nature et les Technologies
(FQRNT) and from the Natural Science and Engineering Research
Council of Canada (NSERC).}

\end{document}